\documentclass[12pt]{iopart}
\usepackage{epsfig}
\begin{document}


\title{A new mathematical model of tunnelling and the Hartman effect puzzle}

\author{N L Chuprikov
\footnote[3]{Also at Physics Department, Tomsk State University} }

\address{Tomsk State Pedagogical University, 634041, Tomsk, Russia}
\begin{abstract}
In this paper we develop and refine our recent model of a one-dimensional completed
scattering, which gives an individual description of the transmission and reflection
subprocesses at all stages of scattering. We show that the group, dwell and Larmor
characteristic times, introduced in this model for the subprocesses, are in a full
agreement with special relativity and allow us to solve the Hartman effect puzzle.
\end{abstract}


\maketitle

\newcommand{\Api}{A^{in}}
\newcommand{\Ami}{B^{in}}
\newcommand{\Apo}{A^{out}}
\newcommand{\Amo}{B^{out}}
\newcommand{\bpi}{a^{in}}
\newcommand{\bmi}{b^{in}}
\newcommand{\bpo}{a^{out}}
\newcommand{\bmo}{b^{out}}
\newcommand{\api}{a^{in}}
\newcommand{\ami}{b^{in}}
\newcommand{\apo}{a^{out}}
\newcommand{\amo}{b^{out}}
\newcommand {\uta} {\tau_{tr}}
\newcommand {\utb} {\tau_{ref}}

\section{Introduction}

For a long time tunnelling a particle through a one-dimensional (1D) static
potential barrier has been considered in quantum mechanics as a representative of
well-understood phenomena. The quantum-mechanical model of this process, hereinafter
referred to as "standard model" (SM), has been included in many textbooks on quantum
mechanics. However, studying the temporal aspects of tunnelling (see reviews
\cite{Ha2,La1,Olk,Ste,Mu0,Nu0,Win} and references therein), on the basis of the SM,
shows that this model leads to anomalously short or even negative tunnelling times.
A serious controversy raised by this result has not been overcome up to now.

Among a huge variety of proposals to solve the tunnelling time problem (TTP), the
concepts of the group, dwell and Larmor times are the most prominent ones. They lie
entirely within the framework of conventional quantum theory and complement each
other in timing a scattering particle within the standard setting of this 1D
scattering problem. However, even these concepts, being introduced within the
framework of the SM, both in the case of the (nonrelativistic) Schr\"odinger
equation (see, e.g., \cite{Har,But,Mu6,Wi1,Ol1,So1}) and the (relativistic) Dirac
equation (see, e.g., \cite{Krek,Zhi,Lun}), lead to the Hartman effect to be at
variance with special relativity. As a result, at present there is no consensus in
solving the TTP.

In our opinion such state of affairs is not occasional, for the TTP cannot be, in
principle, solved within the framework of the SM. The main reason is that the
quantum ensemble of particles, at the final stage of a 1D completed scattering,
consists of two subensembles to occupy macroscopically distinct spatial regions.
This fact implies performing two (infinite) identical series of independent
measurements, separately for transmitted particles and separately for reflected
ones.

The partition of the initial quantum ensemble of particles into two macroscopically
distinct parts is crucial for understanding the nature of this quantum scattering
process. By the probability theory (see \cite{Hre} and references therein),
experimental data obtained in two different sets of measurements (identical in
either set) cannot be described by a single (Kolmogorov) probability space. This
means that the only legitimate way of solving the TTP is an individual timing of
either subensemble. Any averaging over the transmitted and reflected subensembles,
contrary to probability theory, leads inevitably to nonphysical results.

It is evident that the SM does not support an individual timing of the subensembles
in the barrier region. Indeed, such timing needs the knowledge of the whole time
evolution of either subensemble. However, the SM does not obey this requirement.
Thus, on the basis of this model, neither common characteristic times, nor
individual ones can be introduced for the subensembles.

At the same time, as was shown in \cite{Ch26,Ch27}, the Schr\"odinger equation, in
reality, admits a separate description of the subensembles at all stages of
scattering, and hence it is possible to introduce characteristic times for each of
them. The problem, however, is that some aspects of the subensemble's evolution have
remained beyond the scope of these papers. Our aim is just to complete the model
\cite{Ch26,Ch27} and to resolve on this basis the controversy surrounding the
Hartman effect. In doing so, we shall dwell shortly on the basic points of the
model, to make the present paper all-sufficient.

\newcommand{\ko}{\kappa_0^2}
\newcommand{\kj}{\kappa_j^2}
\newcommand{\kd}{\kappa_j d_j}
\newcommand{\kki}{\kappa_0\kappa_j}

\newcommand{\Ra}{R_{j+1}}
\newcommand{\Rb}{R_{(1,j)}}
\newcommand{\Rc}{R_{(1,j+1)}}

\newcommand{\Ta}{T_{j+1}}
\newcommand{\Tb}{T_{(1,j)}}
\newcommand{\Tc}{T_{(1,j+1)}}

\newcommand{\Wa}{w_{j+1}}
\newcommand{\Wb}{w_{(1,j)}}
\newcommand{\Wc}{w_{(1,j+1)}}

\newcommand{\UU}{u^{(+)}_{(1,j)}}
\newcommand{\VV}{u^{(-)}_{(1,j)}}

\newcommand{\ta}{t_{j+1}}
\newcommand{\tb}{t_{(1,j)}}
\newcommand{\tc}{t_{(1,j+1)}}

\newcommand{\tee}{\vartheta_{(1,j)}}

\newcommand{\tta}{\tau_{j+1}}
\newcommand{\ttb}{\tau_{(1,j)}}
\newcommand{\ttc}{\tau_{(1,j+1)}}

\newcommand{\FF}{\chi_{(1,j)}}
\newcommand {\aro}{(k)}
\newcommand {\da}{\partial}
\newcommand{\ppp}{\mbox{\hspace{5mm}}}
\newcommand{\ooo}{\mbox{\hspace{3mm}}}
\newcommand{\ooa}{\mbox{\hspace{1mm}}}
\newcommand{\ppa}{\mbox{\hspace{2cm}}}

\section{Wave functions for the subprocesses of a 1D completed scattering} \label{a1}

Remind that a 1D completed scattering was considered in \cite{Ch26} in the following
setting. A particle impinges a symmetrical potential barrier $V(x)$
($V(x-x_c)=V(x_c-x)$) confined to the finite spatial interval $[a,b]$ $(a>0)$;
$d=b-a$ is the barrier width, the point $x_c$ is the centre of the barrier region.
At the initial instant of time, long before the scattering event, the state of a
particle $\psi_{full}^{(0)}(x)$ approaches the in-asymptote $\psi_{full}^{in}(x,t)$,
\[\fl\psi_{full}^{in}(x,t)=\frac{1}{\sqrt{2\pi}}\int_{-\infty}^{\infty}\Api(k)\exp[i(kx-
E(k)t/\hbar)]dk,\] which is supposed to be a normalized function to belong to the
set $S_{\infty}$ consisting from infinitely differentiable functions vanishing
exponentially in the limit $|x|\to \infty$; $E(k)=\hbar^2k^2/2m$. Without loss of
generality, it is also supposed that
\begin{eqnarray} \label{444}
\fl<\psi_{full}^{(0)}|\hat{x}|\psi_{full}^{(0)}>=0,\ooo
<\psi_{full}^{(0)}|\hat{p}|\psi_{full}^{(0)}> =\hbar k_0
> 0,\ooo <\psi_{full}^{(0)}|\hat{x}^2|\psi_{full}^{(0)}> =l_0^2,
\end{eqnarray}
where $l_0$ and $k_0$ are given parameteres ($l_0<<a$); $\hat{x}$ and $\hat{p}$ are
the operators of the particle's position and momentum, respectively. For the
Gaussian wave packet $\Api(k)=(l_0^2/\pi)^{1/4} \exp(-l_0^2(k-k_0)^2)$. For a
completed scattering the average velocity, $\hbar k_0/m,$ is supposed to be much
more than the rate of spreading the incident wave packet.

For each value of time $t$ the state of a particle has the form
\begin{eqnarray} \label{11}
\fl\psi_{full}(x,t)=\frac{1}{\sqrt{2\pi}}\int_{-\infty}^{\infty}
\Api(k)\psi_{full}(x;k)\exp[-i E(k)t/\hbar];
\end{eqnarray}
$\psi_{full}(x;k)$ describes the stationary state of a particle, which is presented
in \cite{Ch26} as follows
\begin{eqnarray} \label{511}
\fl \psi_{full}(x;k)=\left\{ \begin{array}{c} e^{ikx}+b_{out}(k)e^{ik(2a-x)},\ppa\ppp\ooo for \ooa x\le a;\\
a_{full}\cdot u(x-x_c;k)+b_{full}\cdot v(x-x_c;k), \ppp \ooa for\ooa a\le x\le b;\\
a_{out}(k)e^{ik(x-d)}, \ppa\ppa for \ooa x>b;
\end{array}\right.
\end{eqnarray}
$u(x-x_c;k)$ and $v(x-x_c;k)$ are such real solutions to the Schr\"odinger equation
that $u(x_c-x;k)=-u(x-x_c;k)$, $v(x_c-x;k)=v(x-x_c;k)$ and
$\frac{du}{dx}v-\frac{dv}{dx}u=\kappa$ is a constant;
\begin{eqnarray} \label{51300}
\fl a_{out}=\frac{1}{2}\left(\frac{Q}{Q^*}-\frac{P}{P^*}\right);\ooo
b_{out}=-\frac{1}{2}\left(\frac{Q}{Q^*}+\frac{P}{P^*}\right).
\end{eqnarray}
\begin{eqnarray*}
\fl a_{full}=\frac{1}{\kappa}\left(P+P^*b_{out}\right)e^{ika}=
-\frac{1}{\kappa}P^*a_{out}e^{ika};\ooo
b_{full}=\frac{1}{\kappa}\left(Q+Q^*b_{out}\right)e^{ika}=
\frac{1}{\kappa}Q^*a_{out}e^{ika};\nonumber
\end{eqnarray*}
\begin{eqnarray*}
\fl Q=\left(\frac{du(x-x_c)}{dx}+i k u(x-x_c)\right)\Bigg|_{x=b};\ooa
P=\left(\frac{dv(x-x_c)}{dx}+i k v(x-x_c)\right)\Bigg|_{x=b}.\nonumber
\end{eqnarray*}

As is shown in \cite{Ch26}, the wave function $\psi_{full}(x;k)$ to describe the
stationary state of the whole ensemble of scattering particles can be uniquely
presented as the superposition of the functions $\psi_{tr}(x;k)$ and
$\psi_{ref}(x;k)$ to describe the subensembles of transmitted and reflected
particles, respectively - $\psi_{full}(x;k)=\psi_{tr}(x;k)+\psi_{ref}(x;k)$. Here
\begin{eqnarray*}
\fl\psi_{ref}(x;k)=\Api_{ref}e^{ikx}+b_{out}e^{ik(2a-x)},\ooo \psi_{tr}(x;k)=
\Api_{tr}e^{ikx} \ooo for \ooo x\le a;
\end{eqnarray*}
\begin{eqnarray} \label{2}
\fl \begin{array}{c} \psi_{ref}(x;k)=\kappa^{-1}\left(PA^{in}_{ref}+
P^*b_{out}\right)e^{ika}u(x-x_c;k)\\
\psi_{tr}(x;k)=a_{tr}u(x-x_c;k)+b_{full}v(x-x_c;k)
\end{array}\Bigg\} \ooo for \ooo a\leq x\leq
x_c;
\end{eqnarray}
\begin{eqnarray*}
\fl\psi_{ref}(x;k)\equiv 0,\ooo \psi_{tr}(x;k)\equiv \psi_{full}(x;k) \ooo for \ooo
x\ge x_c;
\end{eqnarray*}
\begin{eqnarray*}
\fl\Api_{ref}=b_{out}\left(b^*_{out}-a^*_{out}\right)\equiv
b_{out}^*\left(b_{out}+a_{out}\right);\ooa
\Api_{tr}=a^*_{out}\left(a_{out}+b_{out}\right)\equiv
a_{out}\left(a^*_{out}-b^*_{out}\right)
\end{eqnarray*}
\begin{eqnarray} \label{3}
\fl a_{tr}=\frac{P}{\kappa}A^{in}_{tr}e^{ika}=-\frac{PQ^*}{P^*Q}\cdot a_{full}.
\end{eqnarray}
We have to stress that not only $\Api_{tr}+\Api_{ref}=1$ but also
$|\Api_{tr}|^2+|\Api_{ref}|^2=1.$ The amplitudes $\Api_{tr}$ and $\Api_{ref}$ can
also be presented in terms of the transmission and reflection coefficients -
$\Api_{ref}=\sqrt{R}(\sqrt{R}\pm i\sqrt{T}) \equiv \sqrt{R}\exp(i\lambda)$,
$\Api_{tr}=\sqrt{T}(\sqrt{T}\mp i\sqrt{R}) \equiv
\sqrt{T}\exp\left[i\left(\lambda+sign(\lambda)\frac{\pi}{2}\right)\right]$;
$\lambda=\pm\arctan(\sqrt{T/R})$; $T=|a_{out}|^2$, $R=|b_{out}|^2$.

The main peculiarity of $\psi_{tr}(x;k)$ and $\psi_{ref}(x;k)$ is that each of them,
unlike $\psi_{full}(x;k)$, contains one incoming and one outgoing wave. As is seen
from (\ref{2}), the unitary (Schr\"odinger's) character of transmission and
reflection is violated at the point $x_c$. At the same time both the functions as
well as the corresponding probability current densities are continuous everywhere on
the $OX$-axis, including the point $x_c$.

By our approach, the point $x_c$ of any symmetrical potential barrier is a special
one. In particular, reflected particles never cross this point in the course of
scattering. This result agrees entirely with the fact that, for classical particles
to impinge from the left a smooth symmetrical potential barrier, the middle of the
barrier region is the extreme right turning point, irrespective of the particle's
mass and the barrier's form and size.

For narrow in $k$-space wave packets $\psi_{full}(x,t)$ (see (\ref{11})) and
corresponding ones $\psi_{tr}(x,t)$ and $\psi_{ref}(x,t)$ formed from
$\psi_{tr}(x;k)$ and $\psi_{ref}(x;k)$, respectively, we have
$\Re\langle\psi_{tr}(x,t)|\psi_{ref}(x,t)\rangle=0$ for any value of $t$. Therefore,
despite the existence of interference between $\psi_{tr}$ and $\psi_{ref}$, we have
\begin{eqnarray*}
\fl \langle\psi_{full}(x,t)|\psi_{full}(x,t)\rangle =\textbf{T}+\textbf{R}=1;
\textbf{T}=\langle\psi_{tr}(x,t)|\psi_{tr}(x,t)\rangle,\ooa
\textbf{R}=\langle\psi_{ref}(x,t)|\psi_{ref}(x,t)\rangle;
\end{eqnarray*}
constants $\textbf{T}$ and $\textbf{R}$ are the transmission and reflection
coefficients, respectively.

Note that the decomposition $\psi_{full}(x,t)=\psi_{tr}(x,t)+\psi_{ref}(x,t)$ holds
for wave packets of any width. In this case $\textbf{R}$ remains unchanged at all
stages of scattering. However, $\textbf{T}$ is now constant and equal to
$1-\textbf{R}$ only long before and long after the scattering event. At the very
stage of scattering this quantity is not now constant: as the wave packet
$\psi_{tr}(x,t)$ does not obey the Schr\"odinger equation at the point $x_c$, the
continuity, at this point, of the probability current density (PCD) of separate
waves does not guarantee the continuity of the PCD for their superposition, for the
continuity equation is nonlinear.

Thus, in the general case, in partitioning the whole ensemble of scattering
particles into the to-be-transmitted and to-be-reflected subensembles at the stage
of scattering, the quantum mechanical formalism does not allow one to exclude
entirely interference terms from $\psi_{tr}(x,t)$. However, as it follows from our
numerical calculations, even for wave packets whose initial width is comparable with
the barrier width, the relative deviation of the value of $\textbf{T}$ from
$1-\textbf{R}$ is small enough. This is a consequence of a large rate of spreading
such packets. At the very stage of scattering the width of such packets becomes much
larger than the barrier's width, which results in weakening the effect of the
violation of the continuity equation at the point $x_c$.

Note that the question of violating the unitary evolution of the subensembles at the
point $x_c$ has remained the scope of the papers \cite{Ch26,Ch27}. To cover this
gap, in the context of solving the Hartman effect puzzle, is the main goal of the
present paper. Of interest here is the fact that due to non-unitarity the time
derivative of the expectation values of observables involved in the timing
procedures of the subensembles may contain extra terms, apart from the quantum
Poisson brackets. To elucidate this question, it is sufficient, for the first time,
to restrict oneself to the case of narrow in $k$-space wave packets when the
variation of $\textbf{T}$ is negligible.

For example, it is easy to show that for such packets reflected electrons are
affected, at the point $x_c$, by an extra (average) force to push particles out from
the barrier region, backward into the left out-of-barrier one -
\[\fl \frac{d<\hat{p}>_{ref}}{dt}=\left<-\frac{dV}{dx}\right>_{ref}-
\frac{\hbar^2}{2m}\left|\frac{\partial\psi_{ref}}{\partial x}\right|^2_{x=x_c-0};\]
where angle brackets denote expected values of observables. For transmitted
particles the second term in the analogous expression
\[\fl \frac{d<\hat{p}>_{tr}}{dt}=\left<-\frac{dV}{dx}\right>_{tr}+
\frac{\hbar^2}{2m}\left(\left|\frac{\partial\psi_{tr}}{\partial
x}\right|^2_{x=x_c+0}-\left|\frac{\partial\psi_{tr}}{\partial
x}\right|^2_{x=x_c-0}\right)\] equals to zero. Indeed, in the limit $l_0\to\infty$
\[\fl \left|\frac{\partial\psi_{tr}}{\partial
x}\right|^2_{x=x_c+0}-\left|\frac{\partial\psi_{tr}}{\partial
x}\right|^2_{x=x_c-0}=\kappa^2\left(|a_{full}|^2-|a_{tr}|^2\right)=0,\] because the
modules of the coefficients $a_{full}(k)$ and $a_{tr}(k)$ are equal (see (\ref{3})).

What is important is that the violation of the unitary subensemble's evolution leads
also to extra terms in the time derivatives for the $x$-th and $y$-th projections of
the electron spin. They are these observables that are used for introducing the
Larmor characteristic times for the subensembles (see \cite{Ch27}). In doing so,
extra terms associated with the non-unitarity have not been considered in
\cite{Ch27} because they do not describe the Larmor spin precession in a magnetic
field confined to the barrier region - an effect to underlie the Larmor timing
procedure. At the same time, as will be seen from the following, the appearance of
such terms plays the key role in solving the old mystery associated with the Hartman
effect.

As is known, the essence of this effect is that, for a particle tunnelling through a
wide rectangular barrier, the phase (asymptotic group) time (see \cite{Smi}) and the
Larmor time to coincide with the dwell time (see \cite{But}) saturate with
increasing the barrier's width. Thus, in fact, to study all aspects of the Hartman
effect, we have to dwell on all characteristic times introduced in \cite{Ch27} for
transmission, taking now into account a non-unitary character of this subprocess.

\section{The Hartman effect puzzle}
\subsection{The Hartman effect from the viewpoint of the group time concept}

We begin our analysis of the Hartman effect with the group time concept. A new model
implies introduction of two different group times - the exact group time
$\tau_{tr}^{exact}$ and the asymptotic group time $\tau_{tr}^{as}$. By \cite{Ch27},
the former is introduced as the difference $\tau_{tr}^{exact}=t^{tr}_2-t^{tr}_1$
where $t^{tr}_1$ and $t^{tr}_2$ are such moments of time that
\begin{eqnarray*}
\fl
\frac{1}{\textbf{T}}\left(<\psi_{tr}(x,t^{tr}_1)|\hat{x}|\psi_{tr}(x,t^{tr}_1)>\right)=a;\ooo
\frac{1}{\textbf{T}}\left(<\psi_{tr}(x,t^{tr}_2)|\hat{x}|\psi_{tr}(x,t^{tr}_2)>\right)=b.
\end{eqnarray*}
As regards $\tau_{tr}^{as}$, it describes the influence of the potential barrier on
a particle within a wide enough interval $[a-L_1,b+L_2]$ where $L_1,L_2\gg l_0.$ In
this case, instead of the exact wave functions for transmission, one may use the
corresponding in- and out-asymptotes
\begin{eqnarray} \label{220}
\fl \psi_{tr}^{in,out}(x,t)=\frac{1}{\sqrt{2\pi}}\int_{-\infty}^{\infty}
\Api(k)f_{tr}^{in,out}(k)\exp[i(kx-E(k)t/\hbar)];
\end{eqnarray}
\begin{eqnarray*}
\fl f_{tr}^{in}(k,t)=\sqrt{T}\exp\left[i\left(\lambda
+sign(\lambda)\frac{\pi}{2}\right)\right],\ooa
f^{out}_{tr}(k)=\sqrt{T}\exp[i(J(k)-kd)];\ooa J=\arg(a_{out}).
\end{eqnarray*}

Long before and long after the scattering event the motion of the centre of mass
(CM) $<\hat{x}>_{tr}$ of the wave packet $\psi_{tr}(x,t)$ is described,
respectively, by the expressions
\begin{eqnarray*}
\fl <\hat{x}>_{tr}^{in}=\frac{\hbar t}{m}<k>_{tr}-<\lambda^\prime>_{tr}^{in};\ooa
<\hat{x}>_{tr}^{out}=\frac{\hbar t}{m}<k>_{tr}-<J^\prime>_{tr}^{out}+d
\end{eqnarray*}
For the average starting point $x^{start}_{tr}$ of transmitted particles we have
$x^{start}_{tr}=-<\lambda^\prime>^{tr}_{in}$, i.e., it differs from
$x^{start}_{full}$ to characterize the whole ensemble of particles. This result
distinguishes our approach from the standard wave-packet analysis based on the
implicit assumption that transmitted (and reflected) particles start, on the
average, from the point $x^{start}_{full}$ to coincide, by setting (\ref{444}), with
the origin of coordinates.

The time $\uta(L_1,L_2)$ spent by the CM $<\hat{x}>_{tr}$ in the interval
$[a-L_1,b+L_2]$ is
\begin{eqnarray*}
\fl \uta(L_1,L_2)\equiv t_{tr}^{(2)}-t_{tr}^{(1)}=\frac{m}{\hbar
<k>_{tr}}\left(<J^\prime>_{tr}^{out} -<\lambda^\prime>_{tr}^{in} +L_1+L_2 \right).
\end{eqnarray*}
The values of $t_{tr}^{(2)}$ and $t_{tr}^{(1)}$ obey the equations
\[\fl <\hat{x}>_{tr}^{in}(t_{tr}^{(1)})=a-L_1;\ooo <\hat{x}>_{tr}^{out}(t_{tr}^{(2)})=b+L_2.
\]

The term $\uta^{as}$ ($\uta^{as}=\uta(0,0)$) is just the asymptotic group
transmission time,
\begin{eqnarray} \label{230}
\fl \uta^{as}=\frac{m d^{gr}_{tr}}{\hbar <k>_{tr}},\ppp
d^{gr}_{tr}=<J^\prime>_{tr}^{out} -<\lambda^\prime>_{tr}^{in}.
\end{eqnarray}
For a particle tunnelling through the rectangular potential barrier of height $V_0$
($E\leq V_0$), with the notations $\kappa=\sqrt{2m(V_0-E)}/\hbar$ and
$\kappa_0=\sqrt{2mV_0}/\hbar$, we have (see \cite{Ch27})
\[\fl d_{tr}^{gr}(k)=\frac{4}{\kappa}
\frac{\left[k^2+\kappa_0^2\sinh^2\left(\kappa d/2\right)\right]
\left[\kappa_0^2\sinh(\kappa d)-k^2 \kappa d\right]} {4k^2\kappa^2+
\kappa_0^4\sinh^2(\kappa d)}.\]

\begin{figure}[h]
\begin{center}
\includegraphics[width=8.0cm]{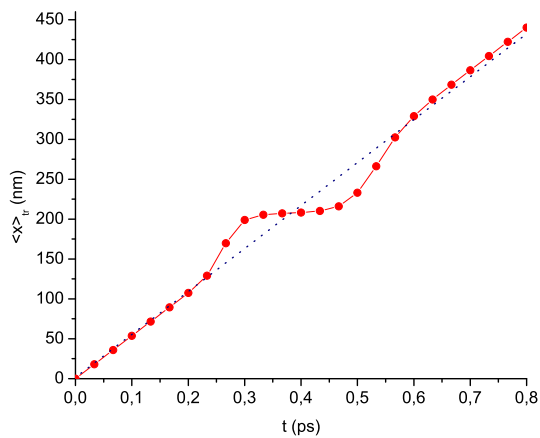}
\end{center}
\caption{The CM's positions of $\psi_{tr}(x,t)$ (circles) and the corresponding
freely moving wave packet (dashed line) as functions of time $t$.}
\label{fig:fig5a1}
\end{figure}

As is seen, like the phase time defined in the SM, this quantity saturates, too,
with increasing the barrier's width $d$. However, this fact does not at all mean
that the effective velocity of a particle tunnelling through a wide rectangular
barrier becomes superluminal. The figure \ref{fig:fig5a1} shows the function
$<\hat{x}>_{tr}(t)$ to describe scattering the Gaussian wave packet ($l_0=10nm$,
$E_0=\hbar^2k_0^2/2m=0.05eV$) by the rectangular barrier ($a=200nm$, $b=215nm$,
$V_0=0.2eV$). (Note, in this case the deviation of $\textbf{T}$ from $1-\textbf{R}$
has not exceeded five percentages, though the wave-packet's and barrier's widths are
of the same order.)

This figure shows explicitly the principal difference between the exact and
asymptotic group times. While the former gives the time spent by the wave packet's
CM just in the barrier region, the latter describes the influence of the barrier on
the CM, in the course of a whole scattering process. More precisely, the quantity
$\uta^{as}-\tau_{free}$, where $\tau_{free}=md/\hbar k_0$, is the time delay in
moving the CM of the transmitted wave packet in comparison with the motion of the CM
of the corresponding freely moving one. In the case considered, $\uta^{exact}\approx
0,155ps$, $\uta^{as}\approx 0,01ps$, $\tau_{free}\approx 0,025ps$.

As is seen, the influence of this opaque rectangular barrier on the transmitted wave
packet has a complicated character. The exact group time says that the CM's velocity
becomes very small inside the barrier region. However, the asymptotic group time
tells us that the total influence of the barrier on the wave packet has an
accelerating character: the wave packet transmitted through the barrier moves ahead
the corresponding freely moving one. However, this effect is related to the
asymptotically large spatial interval. Therefore the saturation of the asymptotic
group transmission time, with increasing the barrier's width, does not at all mean
that the CM of the transmitted wave packet crosses the barrier with a superluminal
effective velocity.

\subsection{The Hartman effect from the viewpoint of the dwell and Larmor time
concepts}\label{a32}

Now we have to analyze the Hartman effect on the basis of the dwell and Larmor time
concepts which are closely connected with each other. Remind that the dwell time for
transmission ($\tau_{tr}^{dwell}$) is introduced in \cite{Ch27} for the stationary
case -
\begin{eqnarray} \label{4005}
\fl \tau_{tr}^{dwell}=\frac{1}{I_{tr}(k)}\int_a^b|\psi_{tr}(x;k)|^2 dx
=\frac{2}{I_{full}(k)}\int_{x_c}^b|\psi_{full}(x;k)|^2 dx
\end{eqnarray}
where $I_{tr}=I_{full}=T(k)\hbar k/m$ is the probability current density; the second
expression in (\ref{4005}) reflects the properties of symmetric potential barriers.
For a particle tunnelling through the rectangular barrier we have (see \cite{Ch27})
\begin{eqnarray} \label{4007}
\fl \tau_{tr}^{dwell}=\frac{m}{2\hbar
k\kappa^3}\left[\left(\kappa^2-k^2\right)\kappa d +\kappa_0^2 \sinh(\kappa
d)\right].
\end{eqnarray}
In the limit $d\to\infty$, this quantity increases exponentially rather than
saturates. Thus, both the exact group time and the dwell time tell us, contrary to
the SM, that the opaque barrier strongly delays the motion of a particle when it
enters the barrier region.

However, the intrigue is that the Larmor transmission time, being the average value
of the dwell time, increases in this limit too \cite{Ch27}. This result seems to
contradict the well-established fact to say that the final readings of the Larmor
clock saturate in this case (see \cite{But}). Of importance is that these
(asymptotic) readings are introduced in terms of the transmitted wave packet and
hence must be the same both in the standard and new models of a 1D completed
scattering.

So that, finding the exponential increase of the dwell and Larmor transmission
times, in the above limit, is an important but not definitive step in solving the
Hartman effect puzzle. One has also to resolve the above discrepancy which remained
unexplained in \cite{Ch27}. To complete solving this puzzle is just the main aim of
the present paper, and, as will be seen from the following, the above non-unitary
character of the time evolution of subprocesses plays a crucial role in this case.
To show this, we reconsider some details of the Larmor-clock procedure introduced in
\cite{Ch27}.

By this procedure, there is a mixture of two electron's ensembles - one of them
consists from electrons with spin to be parallel to the OZ-axis, another is formed
from particles with antiparallel spin - which impinge the potential barrier with a
small constant uniform magnetic field $B$ switched on, in the barrier region, along
the OZ-axis. At any instant of time $t$ the state of the mixture is described by the
spinor $\Psi_{full}(x,t)$ to approach at $t=0$ the in-asymptote
$\Psi_{full}^{in}(x)$ -
\begin{eqnarray} \label{9001}
\fl \Psi_{full}(x,t)=\frac{1}{\sqrt{2}}\left(\begin{array}{c} \psi_{full}^{(\uparrow)}(x,t) \\
\psi_{full}^{(\downarrow)}(x,t) \end{array} \right), \ooo
\Psi_{full}^{in}(x)=\frac{1}{\sqrt{2}}\left(\begin{array}{c} 1 \\ 1
\end{array} \right)\psi_{full}^{in}(x);
\end{eqnarray}
$\psi_{full}^{in}(x)$ is a normalized wave function to satisfy conditions
(\ref{444}) where $l_0$ is large enough.

For electrons with spin up (down), the barrier's height is effectively decreased
(increased) by the value $\hbar\omega_L/2$; where $\omega_L=2\mu B/\hbar$ is the
frequency of the Larmor precession; $\mu$ is the magnetic moment. The corresponding
Hamiltonian has the form
\begin{eqnarray} \label{900200}
\fl \hat{H}=\frac{\hat{p}^2}{2m}+V(x)-\frac{\hbar\omega_L}{2}\sigma_z, \ooo if\ooo
x\in[a,b];\ooo \hat{H}=\frac{\hat{p}^2}{2m}, \ooo otherwise
\end{eqnarray}
hereinafter, $\sigma_x,$ $\sigma_y$ and $\sigma_z$ are the Pauli matrices.

By \cite{Ch26}, each component of the spinor $\Psi_{full}(x,t)$ can be uniquely
presented as a coherent superposition of two probability fields to describe
transmission and reflection (we shall suppose that they are known) -
\begin{eqnarray} \label{9003}
\fl \psi_{full}^{(\uparrow)}(x,t)=
\psi_{tr}^{(\uparrow)}(x,t)+\psi_{ref}^{(\uparrow)}(x,t);\ooo
\psi_{full}^{(\downarrow)}(x,t)=
\psi_{tr}^{(\downarrow)}(x,t)+\psi_{ref}^{(\downarrow)}(x,t).
\end{eqnarray}
As a consequence, the same decomposition takes place for spinor (\ref{9001}) -
\begin{eqnarray} \label{9004}
\fl \Psi_{full}(x,t)= \Psi_{tr}(x,t)+\Psi_{ref}(x,t).
\end{eqnarray}
It is important to stress here that
\begin{eqnarray} \label{900100}
\fl
<\psi_{full}^{(\uparrow\downarrow)}(x,t)|\psi_{full}^{(\uparrow\downarrow)}(x,t)>
=T^{(\uparrow\downarrow)}+R^{(\uparrow\downarrow)}=1;
\end{eqnarray}
\begin{eqnarray} \label{900101}
\fl T^{(\uparrow\downarrow)}=<\psi_{tr}^{(\uparrow\downarrow)}(x,t)|
\psi_{tr}^{(\uparrow\downarrow)}(x,t)>;\ooo
R^{(\uparrow\downarrow)}=<\psi_{ref}^{(\uparrow\downarrow)}(x,t)|
\psi_{ref}^{(\uparrow\downarrow)}(x,t)>;
\end{eqnarray}
$T^{(\uparrow\downarrow)}$ and $R^{(\uparrow\downarrow)}$ are the (constant)
transmission and reflection coefficients for particles with spin up $(\uparrow)$ and
down $(\downarrow)$, respectively.

Note that in-state (\ref{9001}) is the engine state of $\sigma_x$ with the
eigenvalue 1 (the average spin of the ensemble of incident particles is oriented
along the $x$-direction). In the course of the scattering process the average spin
of both transmitted and reflected particles will rotate in the plane orthogonal to
the external magnetic field. However, since our main goal is interpreting the
Hartman effect we will not be interested here in the spin's dynamics of reflected
particles.

To study the spin's dynamics, it is convenient to present the average projections
$\hat{S}_x$, $\hat{S}_y$ and $\hat{S}_z$ of the electron spin for the transmitted
subensemble in the form
\begin{eqnarray} \label{519007}
\fl <\hat{S}_x>_{tr}\equiv
\frac{\hbar}{2}\sin(\theta_{tr})\cos(\phi_{tr})=\frac{\hbar}{2\tilde{\textbf{T}}}
\Re(<\psi_{tr}^{(\uparrow)}|\psi_{tr}^{(\downarrow)}>),\nonumber\\
\fl <\hat{S}_y>_{tr}\equiv
\frac{\hbar}{2}\sin(\theta_{tr})\sin(\phi_{tr})=\frac{\hbar}{2\tilde{\textbf{T}}}
\Im(<\psi_{tr}^{(\uparrow)}|\psi_{tr}^{(\downarrow)}>),\\
\fl <\hat{S}_z>_{tr}\equiv
\frac{\hbar}{2}\cos(\theta_{tr})=\frac{\hbar}{4\tilde{\textbf{T}}}
\Big(<\psi_{tr}^{(\uparrow)}|\psi_{tr}^{(\uparrow)}>
-<\psi_{tr}^{(\downarrow)}|\psi_{tr}^{(\downarrow)}>\Big);\nonumber
\end{eqnarray}
$\tilde{\textbf{T}}=(T^{(\uparrow)}+T^{(\downarrow)})/2$; analogous angles are also
introduced for $\Psi_{full}$ and $\Psi_{ref}$.

For the initial condition (\ref{9001}) $\theta_{full}(t)=\theta_{full}^{(0)}=\pi/2$,
$\phi_{full}^{(0)}\equiv\phi_{full}(0)=0$, however
\begin{eqnarray*}
\fl \theta_{tr}^{(0)}=\arccos\left(\frac{T^{(\uparrow)}-
T^{(\downarrow)}}{T^{(\uparrow)}+T^{(\downarrow)}}\right)\neq\frac{\pi}{2};\ooo
\phi_{tr}^{(0)}=\arctan\left(\frac{\Im(<\psi_{tr}^{(\uparrow)}(x,0)|
\psi_{tr}^{(\downarrow)}(x,0)>)}
{\Re(<\psi_{tr}^{(\uparrow)}(x,0)|\psi_{tr}^{(\downarrow)}(x,0)>)}\right)\neq 0.
\end{eqnarray*}

Note, the norm of the narrow in $k$-space wave packets
$\psi_{tr}^{(\uparrow\downarrow)}(x,t)$ is constant in time, therefore, despite a
non-unitary evolution of transmission, $\theta_{tr}(t)\equiv\theta_{tr}^{(0)}$. In
this case
\begin{eqnarray} \label{5190018}
\fl <\hat{S}_z>_{tr}(t)=\frac{\hbar}{2}\cdot\frac{T^{(\uparrow)}-
T^{(\downarrow)}}{T^{(\uparrow)}+T^{(\downarrow)}}.
\end{eqnarray}
That is, this projection is constant, in a full agreement with the fact that the
operator $\hat{S}_z$ commutes with Hamiltonian (\ref{900200}). Thus, by our
approach, unlike the SM (see \cite{But}), the angle $\theta_{tr}(t)$ cannot be used
as a measure of the duration of dwelling an electron in the barrier region. This
angle is nonzero from the very outset of the scattering process and remains
unchanged in time.

So that only the change of $\phi_{tr}(t)$, due to the Larmor precession, can be used
for measuring the time spent, on the average, by transmitted electrons in the
barrier region. However, by our approach, apart from the Larmor precession there are
other two reasons to influence the value of $\phi_{tr}^{end}$ - final readings of
the Larmor clock. One of them has already known - the initial value of the angle
$\phi_{tr}^{(0)}$ is nonzero, unlike $\phi_{full}^{(0)}$. Another reason, as will be
seen from the following, is associated with a non-unitary character of the
transmission subprocess. To study all peculiarities of the Larmor timing procedure
for transmission, let us calculate the derivative $d\phi_{tr}/dt$.

Since $\phi_{tr}=\arctan\left(<\hat{S}_y>_{tr}/<\hat{S}_x>_{tr}\right)$, we have
\begin{eqnarray} \label{51900180}
\fl \frac{d \phi_{tr}}{dt}=\frac{<\hat{S}_x>_{tr}\frac{d
<\hat{S}_y>_{tr}}{dt}\ooa-<\hat{S}_y>_{tr}\frac{d
<\hat{S}_x>_{tr}}{dt}}{<\hat{S}_x>_{tr}^2+<\hat{S}_y>_{tr}^2}.
\end{eqnarray}
Calculations for the derivatives of the corresponding Pauli matrices show that
\begin{eqnarray} \label{51900281}
\fl \tilde{\textbf{T}}\cdot\frac{d<\sigma_x>_{tr}}{dt}=\omega_L \int_a^b
\Im[(\psi_{tr}^{(\uparrow)}(x,t))^*\psi_{tr}^{(\downarrow)}(x,t)]dx-\nonumber\\
\fl -\frac{\hbar}{2m}\Im\Bigg[\psi_{tr}^{(\downarrow)}(x_c,t) \left(\frac{\partial
\psi_{tr}^{(\uparrow)}(x_c+0,t)}{\partial x}- \frac{\partial
\psi_{tr}^{(\uparrow)}(x_c-0,t)}{\partial x}\right)^*- \nonumber\\ - \fl
\psi_{tr}^{(\uparrow)*}(x_c,t) \left(\frac{\partial
\psi_{tr}^{(\downarrow)}(x_c+0,t)}{\partial x}- \frac{\partial
\psi_{tr}^{(\downarrow)}(x_c-0,t)}{\partial x}\right) \Bigg];\\
\fl \tilde{\textbf{T}}\cdot\frac{d<\sigma_y>_{tr}}{dt}=-\omega_L \int_a^b
\Re[(\psi_{tr}^{(\uparrow)}(x,t))^*\psi_{tr}^{(\downarrow)}(x,t)]dx+\nonumber\\
\fl +\frac{\hbar}{2m}\Re\Bigg[\psi_{tr}^{(\downarrow)}(x_c,t) \left(\frac{\partial
\psi_{tr}^{(\uparrow)}(x_c+0,t)}{\partial x}- \frac{\partial
\psi_{tr}^{(\uparrow)}(x_c-0,t)}{\partial x}\right)^*- \nonumber\\ - \fl
\psi_{tr}^{(\uparrow)*}(x_c,t) \left(\frac{\partial
\psi_{tr}^{(\downarrow)}(x_c+0,t)}{\partial x}- \frac{\partial
\psi_{tr}^{(\downarrow)}(x_c-0,t)}{\partial x}\right) \Bigg]\nonumber
\end{eqnarray}
(as $\Psi_{ref}(x_c,t)=0$, for reflection, the second terms in similar expressions
do not appear.)

Let now a magnetic field be infinitesimal. Then, considering (\ref{9001}), we have
\[\fl \left|<\hat{S}_y>_{tr}\right|\ll\left|<\hat{S}_x>_{tr}\right|,\ppp
\left|\frac{d<\sigma_x>_{tr}}{dt}\right|\ll
\left|\frac{d<\sigma_y>_{tr}}{dt}\right|\sim \omega_L.\] So that Exp.
(\ref{51900180}) for $d \phi_{tr}/dt$ becomes simpler - $\frac{d
\phi_{tr}}{dt}=\frac{d<\hat{S}_y>_{tr}}{dt}\Big/<\hat{S}_x>_{tr}$.

As the value of $\omega_L$ is small in (\ref{900200}), the functions
$\psi^{(\uparrow)}(x,t)$ and $\psi^{(\downarrow)}(x,t)$ can be written in the form
$\psi^{(\uparrow)}\approx \psi-\omega_L \tilde{\psi}$, $\psi^{(\downarrow)}\approx
\psi+\omega_L \tilde{\psi}$ where $\psi$ and $\tilde{\psi}$ do not depend on
$\omega_L$. Then, keeping in Exps. (\ref{519007}) and (\ref{51900281}) only the main
terms, we obtain
\begin{eqnarray} \label{519009}
\fl \frac{d \phi_{tr}}{dt}=-\frac{\omega_L}{\textbf{T}} \int_a^b
|\psi_{tr}(x,t)|^2dx- \frac{\hbar \omega_L}{m \textbf{T}}\Re\Bigg[\psi_{tr}(x_c,t)
\left(\frac{\partial \tilde{\psi}^*_{tr}(x_c+0,t)}{\partial x}- \frac{\partial
\tilde{\psi}^*_{tr}(x_c-0,t)}{\partial x}\right)- \nonumber\\\fl -
\tilde{\psi}^*_{tr}(x_c,t) \left(\frac{\partial \psi_{tr}(x_c+0,t)}{\partial x}-
\frac{\partial \psi_{tr}(x_c-0,t)}{\partial x}\right) \Bigg].
\end{eqnarray}

So, there are two reasons that lead to the change of the angle $\phi_{tr}$ in the
course of scattering - the Larmor precession of the average spin of transmitted
electrons in the magnetic field switched on in the barrier region and breaking the
unitary evolution of this subensemble at the point $x=x_c$.

Note that both the terms in (\ref{519009}) are zero long before and long after the
scattering event. So that, to simplify the definition of the total angle
$\Delta\phi_{tr}$ of the spin rotation for a 1D completed scattering, one may shift
the time of beginning and finishing this process to the minus and plus infinity,
respectively. Thus,
\begin{eqnarray} \label{5190020}
\fl \Delta\phi_{tr}=\int_{-\infty}^\infty\frac{d \phi_{tr}}{dt} dt= -\omega_L
\left(\tau^L_{tr}+\tau_{int}\right),
\end{eqnarray}
where $\tau^L_{tr}$ is the Larmor transmission time. Considering Exp. (\ref{519009})
in (\ref{5190020}), we obtain
\begin{eqnarray} \label{5190022}
\fl \tau^L_{tr}=\frac{1}{\textbf{T}} \int_{-\infty}^\infty dt \int_a^b
dx|\psi_{tr}(x,t)|^2.
\end{eqnarray}
So, the larger is the probability of finding a particle in the barrier region, the
larger is the value of $\tau^L_{tr}$. The second term in $\Delta\phi_{tr}$,
associated with a non-unitary evolution of $\psi_{tr}(x,t)$, has no relation to the
average duration of staying a particle in the barrier region -
\begin{eqnarray} \label{51900220}
\fl \tau_{int}=\frac{\hbar}{m
\textbf{T}}\int_{-\infty}^\infty\Re\Bigg[\psi_{tr}(x_c,t) \left(\frac{\partial
\tilde{\psi}^*_{tr}(x_c+0,t)}{\partial x}- \frac{\partial
\tilde{\psi}^*_{tr}(x_c-0,t)}{\partial x}\right)- \nonumber\\ \fl -
\tilde{\psi}^*_{tr}(x_c,t) \left(\frac{\partial \psi_{tr}(x_c+0,t)}{\partial x}-
\frac{\partial \psi_{tr}(x_c-0,t)}{\partial x}\right) \Bigg]dt.
\end{eqnarray}

For Gaussian-like wave packets, the integral over the time interval
$(-\infty,\infty)$, in Exps. (\ref{5190022}) and (\ref{51900220}), can be
calculated. Since
\begin{eqnarray} \label{5190023}
\fl \psi_{tr}(x,t)=\frac{1}{\sqrt{2\pi}}\int_{-\infty}^{\infty}
A^{in}(k)\psi_{tr}(x;k)e^{-i E(k)t/\hbar}dk,
\end{eqnarray}
where $\psi_{tr}(x;k)$ is defined by Exps. (\ref{2}) and (\ref{3})), we have (see
\cite{Ch27})
\begin{eqnarray} \label{5190024}
\fl
\tau^L_{tr}=\frac{1}{\textbf{T}}\int_{0}^{\infty}\varpi(k)T(k)\tau^{tr}_{dwell}(k)dk,
\end{eqnarray}
where $\varpi(k)=|A^{in}(k)|^2-|A^{in}(-k)|^2$; note, for a completed scattering
$|A^{in}(k_0)|\gg|A^{in}(-k_0)|$. Similarly, Exp. (\ref{51900220}) for $\tau_{int}$
is reduced to the form
\begin{eqnarray} \label{51900241}
\fl \tau_{int}=\frac{1}{\textbf{T}}\int_{0}^{\infty}\varpi(k)T(k)\tau_{int}(k)dk;
\end{eqnarray}
\begin{eqnarray} \label{51900240}
\fl \tau_{int}(k)=\frac{1}{k T(k)}\Re\Bigg[\psi_{tr}(x_c;k) \left(\frac{\partial
\tilde{\psi}^*_{tr}(x_c+0;k)}{\partial x}- \frac{\partial
\tilde{\psi}^*_{tr}(x_c-0,k)}{\partial x}\right)-\nonumber \\ \fl -
\tilde{\psi}^*_{tr}(x_c;k) \left(\frac{\partial \psi_{tr}(x_c+0;k)}{\partial x}-
\frac{\partial \psi_{tr}(x_c-0;k)}{\partial x}\right) \Bigg].\nonumber
\end{eqnarray}

As regards the initial and final values of $\phi_{tr}$, then $\phi^{(0)}_{tr}=
-2\omega_L\Im<\tilde{\psi}_{tr}^{in}|\psi_{tr}^{in}>\equiv -\omega_L
\tau^{(0)}_{tr}$, $\phi^{end}_{tr}=
-2\omega_L\Im<\tilde{\psi}_{tr}^{out}|\psi_{tr}^{out}>\equiv -\omega_L
\tau^{end}_{tr}$. So that, in addition to the relationship (\ref{5190020}), the
following expressions must be also true:
$\Delta\phi_{tr}\equiv\phi^{end}_{tr}-\phi^{(0)}_{tr}=-\omega_L
\left(\tau^{end}_{tr}-\tau^{(0)}_{tr}\right)$. Hence, finally, we have
\begin{eqnarray} \label{51900242}
\fl \tau^{end}_{tr}=\tau^{(0)}_{tr}+\tau^L_{tr}+\tau_{int},
\end{eqnarray}
i.e., the final readings $\tau^{end}_{tr}$ of the Larmor clock do not give the time
spent by transmitted particles in the barrier region. Thus, now we can explain the
Hartman effect.

We have to stress once more that the time $\tau^{end}_{tr}$, being defined in terms
of the transmitted wave packet, is the same both in the standard and new models of a
1D completed scattering. For the most interesting case, namely for a particle with
energy $E$, which tunnels through the rectangular barrier ($E<V_0$), we have
\begin{eqnarray} \label{5190029}
\fl \tau^{end}_{tr}(k)= \frac{m k}{\hbar\kappa}\cdot\frac{2\kappa
d(\kappa^2-k^2)+\kappa^2_0\sinh(2\kappa d)}{4k^2\kappa^2+\kappa_0^4\sinh^2(\kappa
d)}
\end{eqnarray}
(by the SM (see \cite{But}), $\tau^{end}_{tr}$ is equal to the dwell time). Besides,
as is shown in \cite{Ch27},
\begin{eqnarray} \label{5190028}
\fl \tau^{(0)}_{tr}(k)= \frac{2mk}{\hbar\kappa}\cdot\frac{(\kappa^2-k^2)\sinh(\kappa
d)+\kappa^2_0\kappa d\cosh(\kappa d)}{4k^2\kappa^2+\kappa_0^4\sinh^2(\kappa d)}
\end{eqnarray}
(it is also useful to note that $\tau^{(0)}_{ref}(k)=\tau^{(0)}_{tr}(k)$,
$\tau^{end}_{ref}(k)=\tau^{end}_{tr}(k)$). As regards $\tau^L_{tr}$, as it follows
from Exp. (\ref{5190024}), in the limit $l_0\to\infty$ considered here, it coincides
with the dwell time $\tau^{tr}_{dwell}(k)$.

So, as it follows from Exp. (\ref{5190029}), $\tau^{end}_{tr}(k)$ does saturate with
increasing the barrier's width. Within the SM where this quantity gives directly the
time spent by a particle in the barrier region, this result leads to the
contradiction with special relativity. However, in our model we meet a principally
different situation. Now the tunnelling time increases exponentially with the
increasing of $d$, so that the effective velocity of electrons to enter the barrier
region decreases exponentially rather than increases beyond all bounds, as it
follows from the SM. As regards $\tau^{end}_{tr}(k)$, for wide rectangular barriers
this quantity is small due to the term $\tau_{int}$ to be negative by value and
comparable with $\tau^{dwell}_{tr}(k)$. The role of the initial readings
$\tau^{(0)}_{tr}(k)$ is not so essential because, in the case considered,
$|\tau^{(0)}_{tr}|\ll\tau^{end}_{tr}$.

\section{Conclusion}

So, the fact of the saturation of the asymptotic group time and the final readings
of the Larmor clock, with increasing the barrier's width, for electrons tunnelling
trough a wide rectangular barrier, does not at all mean that the effective velocity
of tunnelling can be superluminal. As it follows from our model, none of these
characteristic times gives the time spent, on the average, by transmitted electrons
in the barrier region. In our model, the latter is described by the exact group
transmission time and the dwell transmission time. Both these quantities show that
the effective velocity of an electron decreases exponentially when it enters the
region of a wide rectangular barrier.

\section{Acknowledgments}

The author expresses his gratitude to the Programm of supporting the leading
scientific schools of RF (grant No 2553.2008.2) for partial support of this work.

\end{document}